\documentclass[aps,pra,superscriptaddress,preprint,amsmath,amssymb]{revtex4}
\usepackage{epstopdf}
\usepackage{graphicx}
\usepackage{amsmath}
\usepackage[french,english]{babel}
\usepackage{natbib}

\begin{document}
\title{Application of B-splines to determining eigen-spectrum of Feshbach molecules }
\author{A. Derevianko}
\address{Physics Department, University of Nevada, Reno, Nevada  89557, USA}
\author{E. Luc-Koenig}
\address{
Laboratoire Aime Cotton,
Bat. 505, Campus d'Orsay,
91405 ORSAY Cedex, France}
\author{F. Masnou-Seeuws}
\address{
Laboratoire Aime Cotton,
Bat. 505, Campus d'Orsay,
91405 ORSAY Cedex, France}

\date{\today}
\begin{abstract}
The B-spline basis set method is applied to determining the rovibrational eigen-spectrum of
diatomic molecules. A particular attention is paid to a challenging numerical
 task of an accurate and efficient description
of the vibrational levels near the dissociation limit (halo-state and Feshbach
molecules).  Advantages of using B-splines are highlighted by comparing
the performance of the method with that of the commonly-used discrete variable
representation (DVR) approach. Several model cases, including the Morse potential
and realistic potentials with $1/R^3$ and $1/R^6$ long-range dependence of the
internuclear separation are studied.
We find that the B-spline method is superior to the DVR approach and it
is robust enough to properly describe the Feshbach molecules. The developed
numerical method is applied to studying the universal
relation of the energy of the last bound state to the scattering length.
We numerically illustrate the validity of the
quantum-defect-theoretic formulation of such a relation for a $1/R^6$ potential.
\end{abstract}
\maketitle
\section{Introduction}

Finite basis sets technique is an important numerical tool in solving
quantum mechanical problems, e.g., in quantum chemistry~\cite{DavFel86}. One
of the popular recent developments is the use of B-splines in such
calculations. In atomic physics, the applications of B-splines
were stimulated by  Walter R. Johnson's work~\cite{JohBluSap88} and here,
in
this special issue dedicated to celebrating his contributions to atomic physics,
we are delighted to present yet another robust application of B-splines.

The reason to the popularity of the B-splines in practical applications
is due to the fact that  they  form a sufficiently complete basis set with a
reasonably small number of basis functions. Numerical accuracy of the
calculations approaches that of the traditional finite-difference methods,
such as the Numerov method~\cite{BacCorDec01}, with the advantage of a global non-iterative  determination of eigenergies and eigenstates.

Here we apply the B-spline method to study rovibrational eigen-spectrum of
diatomic molecules and  compare
the performance of the method with that of the discrete variable
representation (DVR) approach. Previously, the B-spline method was
successfully applied to finding vibrational spectrum of the Morse potential
in Refs.~\cite{BacCorDec01,Sho73}. Here we focus on the  more challenging problem
of describing vibrational states near the dissociation limit of realistic
long-range potentials. One difficulty lies in the variation of the local de Broglie wavelength by several orders of magnitude from the short range to the long range region. Several authors \cite{kokoouline1999,WilDulMas04} have discussed how the efficiency of the DVR methods could be improved via the implementation of a mapping procedure, where the grid step is adapted to the variation of the de Broglie wavelength. In the present paper we compare the mapped sine grid method of Ref. \cite{WilDulMas04} to  B-splines calculations also using a mapping procedure.
Molecular bound states near dissociation limit play an important
role in the formation of ultracold molecules \cite{masnou2001, doyle2004} and in the determination of   scattering properties in the low-energy regime,  in particular
scattering length \cite{crubellier99} or more generally the threshold energy-dependence of the phaseshifts. The vibrational wavefunctions then extend to  distances much larger than the typical length of the chemical
bond.
Recently several experimental groups succeeded in making loosely-bound ultracold molecules
by sweeping B-fields through the magnetically-induced Feshbach resonances
(see review~\cite{KohGorJul06}).
Such Feshbach molecules may be considered as halo-state systems, since the vibrational wavefunctions extend well into the classi\-cally-for\-bidden
region.  While the halo-state systems, due to their universal behavior for
a wide range of quantum-mechanical systems, deserve a special attention on their own
right~\cite{JenRiiFed04}, there are emerging applications based on the Feshbach molecules: for example,
several schemes of transferring Feshbach molecules to lower vibrational
levels and down to $v=0$ \cite{Koch2004,PerShaSto07} have been proposed.
In the prerequisite numerical time--dependent studies, an expansion
over a suitably-chosen quasi-spectrum is required, and the initial state near dissociation limit
has to be well represented by this quasi--spectrum. The challenge there is the accurate representation of the evanescent part of the wavefunction in the non classical region. B-splines,
with their superior numerical performance, demonstrated here,
may prove useful in such theoretical studies. {We shall therefore evaluate this performance by comparing to analytical results when available (bound levels of the Morse potential)  or to well--established numerical methods.

Motivated by the
spectacular developments in low-energy collision physics of  ultracold atoms,
the universal laws governing near-threshold physics have generated a considerable
interest over the last decade.
In particular, here, with the developed numerical method,
we  investigate a relation
of the energy of the last bound state
to the scattering length.
For potentials without a long range tail, such a  relation is a well-known prediction of the effective-range theory (see e.g.,~\cite{LanLif97}).
For the van der Waals potentials,  the effective range theory has to be improved to account for their asymptotic behavior  and Gao~\cite{Gao04a} has recently derived the proper law relating these two
quantities in the framework of the quantum defect theory (QDT).
Here, using the developed B-spline code, we verify numerically the validity
of this new formulation. We find that compared
to the effective-range result, the  QDT expression remains
accurate over a much wider range of parameters than expected.

The paper is organized as follows. First, in Section~\ref{Sec:Setup},
we set-up the numerical method using the Galerkin technique and expansion
of the molecular wavefunctions over the B-spline basis. We also describe an efficient
molecular grid used in the calculations and recapitulate main features of the DVR method.
In Section~\ref{Sec:NumericalExamples}, we apply the method to finding
ro-vibrational spectra of various potentials and  compare the results
with those from the
DVR method.  We start with the Morse potential,
where analytical results are available, and proceed to realistic potentials, varying
with the internuclear separations, $R$, as $1/R^3$ and $1/R^6$ at large $R$.
With the developed method, in Section~\ref{Sec:eff-range},
we analyze the relation between the scattering length and the position
of the last bound state and compare our numerical results with the predictions
of the QDT and the effective-range theories.
Finally, the conclusions are drawn in Section~\ref{Sec:Conclusion}.

\section{Problem setup}
\label{Sec:Setup}
We are interested in solving the radial time-independent Schr\"{o}\-din\-ger  equation
for vibrational motion of nuclei of a diatomic molecule
\begin{eqnarray}
\nonumber \lefteqn{-\frac{1}{2\mu}u''_{J}\left(  R\right)+
\left(  V\left(
R\right)  +\frac{J\left(  J+1\right)  }{2 \mu R^{2}}\right)  u_{J}\left(  R\right)
=}\\
&&E~u_{J}\left(  R\right)  \, , \label{Eq:DifSchroed}%
\end{eqnarray}
where $\mu$ is the reduced molecular mass, $J$ is the rotational quantum
number and $V\left(  R\right)  $ is the electronic Born-Oppenheimer potential.
Unless specified otherwise, atomic units, $\hbar=\left\vert e\right\vert
=m_{e}\equiv1$, are used throughout.

\subsection{B-spline approach}
General mathematical introduction to B-splines and a collection of codes to
manipulate these basis functions may be found in Ref.~\cite{deB01}. Here we briefly
recapitulate properties of the B-splines relevant to our discussion. We deal with a
set of $n$  functions defined on a support grid $\{t_{i}\}$. A B-spline,
$B_{i}^{\left(  k\right)  }\left(  R\right)  ,$ number $i$ of order $k$ is a
piecewise polynomial of degree $k-1$ inside an interval of the support grid
$t_{i}\leq R<t_{i+k}$.  It vanishes outside this support interval.  The
B-splines are positive functions on their  support interval. In applications,
the common choice (also used here) is to make the end-points of the support grid $k$-fold
degenerate,
\begin{eqnarray*}
t_{1}   & =& t_{2}=\cdots=t_{k}=R_{\min},\\
t_{n+1} & =& \cdots=t_{n+k}=R_{\max},
\end{eqnarray*}
where $n$ is the total number of B-splines in the set. With such a choice of
the grid, the first B-spline, $B_{i=1}^{\left(  k\right)  }\left(  R\right)  $
is the only spline which does not vanish at  $R_{\min}$. Similarly, the only
non-vanishing B-spline at the end-point $R_{\max}$ is the last B-spline,
$B_{i=n}^{\left(  k\right)  }\left(  R\right)$.  Notice that the spline
support grid ${t_i}$ directly maps on the radial grid, except the multiply
defined end-points that map onto the first and last points of the radial grid.

Below, we employ the Galerkin method to obtain a quasi-spectrum of the radial
Schr\"{o}dinger equation, see, e.g., \cite{Joh07book}.
Central to this approach is an
observation that  the differential equation (\ref{Eq:DifSchroed}) may
be derived by seeking an extremum of the action integral,
\begin{eqnarray*}
S&=&\int_{R_{\mathrm{min}}}^{R_{\mathrm{max}}}\left\{  \frac{1}{2\mu}\left(
\frac{du_{J}\left(  R\right)  }{dR}\right)  ^{2}+ \right. \\
&& \left. \left(  V\left(  R\right)
+\frac{J\left(  J+1\right)  }{2 \mu R^{2}}\right)  u_{J}^{2}\left(  R\right)
\right\}  dR \\
&&-E\int_{R_{\mathrm{min}}}^{R_{\mathrm{max}}}u_{J}%
^{2}\left(  R\right)  dR.
\end{eqnarray*}
Further, we expand the ro-vibrational wavefunctions in terms of the B-spline
set,%
\begin{equation}
u_{J}\left(  R\right)  =\sum_{i=2}^{n-1}c_{i}B_{i}^{\left(  k\right)  }\left(
R\right)  .\label{Eq:Expansion}%
\end{equation}
Notice that we discard the first and the last B-spline of the set to enforce
the boundary conditions $u_{J}\left(  R_{\mathrm{min}}\right)  =0$
and $u_{J}\left(  R_{\mathrm{max}}\right)  =0$. The remaining splines vanish
identically at the end-points of the grid.

We substitute the expansion (\ref{Eq:Expansion}) in the action integral and
seek its extremum with respect to the expansion coefficients. As a result, we
arrive at the generalized eigen-value equation for the vector of the
coefficients $\mathbf{c}=(c_{2},c_{3},...c_{n-1})$:
\begin{equation}
\mathbf{Ac}=E~\mathbf{B~c}\,,\label{Eq:genEigenValEq}%
\end{equation}
with matrices
\begin{eqnarray}
A_{ij} &  =&\int_{R_{\mathrm{min}}}^{R_{\mathrm{max}}}\left\{  \frac{1}{\mu
}\frac{dB_{i}}{dR}\frac{dB_{j}}{dR} + \right.  \nonumber \\
&& \left. 2B_{i}\left(  V\left(  R\right)
+\frac{J\left(  J+1\right)  }{2R^{2}}\right)  B_{j}\right\}
dR,\label{Eq:Matrixes}\\
B_{ij} &  =&\int_{R_{\mathrm{min}}}^{R_{\mathrm{max}}}B_{i}B_{j}~dR.\nonumber
\end{eqnarray}
The resulting eigenfunctions are orthonormal and form a numerically complete
basis set in the space of piece-wise polynomials of order $k-1$. The choice of the number of basis functions is determined by the nodal structure of the wavefunctions that we wish to represent.
\subsection{Mapped grid method}
Choosing  numerical grid for solving the radial Schr\"{o}dinger equation for
loosely bound molecules requires special consideration. Realistic potentials
support a large number of bound states. Near the dissociation limit the corresponding  wavefunctions have a large number of nodes. Moreover,the distance between two  nodes, and hence the local De Broglie wavelength,   grows larger as we approach the outer turning point of the
potential. A large fraction of the wavefunction (especially for halo-state
molecules) may reside in the classically-forbidden region. Because of this
behavior of the vibrational states, here we depart from the usual choice of
the radial grid of a constant step as in Refs.~\cite{BacCorDec01,Sho73}.
Instead, we employ a more efficient grid as prescribed by the \textquotedblleft mapped
grid\textquotedblright\ method of  Ref.~\cite{kokoouline1999,WilDulMas04}, first implemented in the framework of the DVR method described below.

In the ``mapped grid'' method
 the radial grid is based on the adaptive coordinate defined as
\[
x\left(  R\right)  =\beta^{-1}\frac{\sqrt{2\mu}}{p_{\max}}\int
_{R_{\mathrm{min}}}^{R}dR^{\prime}\sqrt{E_{\max}-V_{\mathrm{env}}\left(
R^{\prime}\right)  } ,
\]
where $V_{\mathrm{env}}\left(  R\right)  $ is the enveloping potential (it is
chosen to be either the same as or  slightly deeper than the original
potential $V(R)$), $R_{\mathrm{min}}$ is somewhat smaller than the position of
the repulsive inner part of the potential, $E_{\max}$ is the maximum energy
for which accurate results are wanted and $p_{\max}$ is the corresponding value of the
total linear momentum. The grid transformation $x\left(  R\right)  $
efficiently rescales the radial coordinate by the local de Broglie wavelength.
Factor $\beta\leq1$ makes the radial step smaller than the local de Broglie
wavelength and improves the representation of the wavefunction in the
classically-forbidden region. We use a constant step of $\Delta x=\pi
\hbar/p_{\max}$ for the adaptive coordinate. This choice translates into a
variable step of the radial grid,
\begin{equation}
\Delta R\approx\beta\frac{1}{\sqrt{2\mu}}\frac{1}{\sqrt{E_{\max}%
-V_{\mathrm{env}}\left(  R\right)  }}.\label{Eq:RadialStep}%
\end{equation}

At this point we recast the solution of the differential equation in terms of
the generalized eigenvalue equation (\ref{Eq:genEigenValEq}). To
solve this problem, we developed a
numerical code using B-spline routines of Ref.~\cite{deB01}.
Below we evaluate the performance of the method by studying
the rovibrational spectrum of
several potentials.

\subsection{Discrete variable approach}

The DVR approach to the computation of vibrational wavefunctions   \cite{rkosloff1996}, is based on a \emph{collocation} scheme. A wavefunction $\varphi$ is approximated by its projection $\hat P \varphi$ on  a linear combination of $N$ interpolation functions, such that  $\varphi$ and  $\hat P  \varphi$ have the same values at the collocation points. The wavefunction $\varphi$ is thus represented by its values at the collocation points.  The Hamiltonian is represented by a matrix, which can be used to compute bound and continuum states or to simulate the temporal evolution of a wavepacket.   Spectral and collocation methods are discussed in a famous monograph by D. Gottlieb and S. Orszag \cite{dgottlieb1977}.

A great variety of systems have been studied, using various sets of orthogonal interpolation functions. In contrast with the B-splines, such functions do not vanish outside a small interval, but rather  they all are defined on the whole grid, and differ by the number of nodes.

 For applications to ultracold molecules, with  bound and quasi-bound vibrational levels in asymptotically $R^{-6}$ and $R^{-3}$ potentials, Kokoouline {\it et al}\cite{kokoouline1999, kokoouline2000a,kokoouline2000b}   have implemented a  Mapped Fourier Grid method where the interpolation functions are plane waves. The grid step is rescaled  to the value of the local de Broglie wavelength, as described above in Eq.(\ref{Eq:RadialStep}). Accurate results were obtained both for the vibrational energies and for the wavefunctions, using a number of basis functions slightly larger than the number of nodes of the wavefunction of the upper level . The accuracy could be checked by comparison with asymptotic methods \cite{LeRBer70} derived from generalized quantum defect theory .  However, the occurrence   of ghosts levels after diagonalization of the Hamiltonian matrix appeared as a drawback of the mapping procedure.  Willner {\it et al} \cite{WilDulMas04} have shown that when  replacing the plane waves by a basis of $N$ sine functions of the adaptive coordinate  $x$,
\begin{equation}
s_k(x) = \sqrt{\frac{2}{N}}\,\sin\left(k\frac{\pi}{L}x\right)\qquad (k = 1,\dots,N-1),
\end{equation}
 with nodes at both ends of the grid, most of the ghost levels would disappear.\\
 The relevant formulae for the collocation scheme can be found in Ref. \cite{WilDulMas04}.
Note that the number of basis functions is entirely determined by the number of grid steps.
The length of the grid is related to the constant grid step $\delta x$ in the $x$ coordinate by
 \begin{equation}
 L=N \delta x
 \end{equation}
Levels of the Cs$_2$ dimer with a binding energy as small as $\sim 10^{-16}$ a. u. could be computed, for which the vibrational wavefunction extends up to 100 000 a$_0$, i.e. a few tens of microns. This  wavefunction with 528 nodes is computed with a grid of only 706 points:  it is typical of  a halo molecule, most of the probability density lying in the classically forbidden region. The efficiency of a set of oscillating sine functions to represent this slowly decreasing exponential function is then questionable. A discussion on the appearance of ghost levels shows that they are influenced by the value chosen for  the parameter $\beta$ :  a compromise has to be found between the suppression of ghost states ($\beta$ small) and a minimum value of grid points ($\beta \sim 1$). Moreover, the numerical representation of the potential, where an analytical long range behavior is usually matched to an interpolation function between discrete ab initio data at short range may be a source of unphysical levels
Our choice in the present paper is to compare the efficiency of the B-spline and sine-grid  methods for the same grid.

\section{Numerical examples}
\label{Sec:NumericalExamples}
\subsection{Morse potential}
As a test of the quality of our numerical approach, we start with the Morse
potential~\cite{Mor29},  which has no long range tail  but has an advantage of having analytically known
energy levels and wavefunctions. The Morse potential is given by
\begin{equation}
V\left(  r\right) = D\left(  e^{-2a\left(  r-r_{0}\right)  }-2e^{-a\left(
r-r_{0}\right)  }\right)  \,,
\end{equation}
where $D$ is the dissociation energy, $r_{0}$ is the equilibrium position, and
the parameter $a$ governs the spatial extent of the potential. The energies of
the bound states are known exactly,
\begin{equation}
E_{v}=-D+\hbar\omega_{0}\left(  v+1/2\right)  -\left(  \frac{\hbar\omega_{0}%
}{4D}\right)  \hbar\omega_{0}\left(  v+1/2\right)  ^{2},
\end{equation}
where the vibrational quantum number $v=0,1,..v_{D}$, with the maximum,
$v_{D}=\left\lfloor a^{-1}\sqrt{2\mu D}-1/2\right\rfloor $. In these formulas,
the vibrational frequency is
\begin{equation}
\omega_{0}=a\left(  \frac{2D}{\mu}\right)  ^{1/2}\,.
\end{equation}

In calculations we use Morse potential fitted to the ground state potential of
$^{133}$Cs$_{2}$ dimer. The parameters of the employed Morse potential are (in
atomic units) $r_{0}=8.77$, $D=0.016627$, $a=0.372031199$. This potential
supports 170 bound states.

We carry out the DVR and B-spline computations using identical grids.
Given the
same grid, the accuracy of the resulting eigen-values depends only on the
basis, sin (DVR) or B-spline set, and the method of solution of the
Schr\"{o}dinger equation (collocation versus Galerkin method).
In Table~\ref{Tab:Morse}, we compare the computed energies (both DVR and B-splines) with
analytical results for vibrational levels near the dissociation limit. The
results marked $a$ were computed using a relatively small grid of $N=275$
points ($R_{\mathrm{min}}=6.3\,a_{0}$, $R_{\mathrm{max}}=100\,a_{0}$, and
$\beta=0.7$). The larger and denser grid (entries marked $b$) has $N=553$
points, $R_{\mathrm{min}}=6.3\,a_{0}$, $R_{\mathrm{max}}=2000\,a_{0}$, and
$\beta=0.4$. In both cases $E_{\mathrm{max}}=10^{-8}$. The order of B-splines
is $k=15$.

First we consider a case of the coarse grid (a). The accuracy of reproducing the
energies of the low-lying states in the B-spline method is at the level of
$10^{-11}\,\mathrm{cm}^{-1}$, while the DVR method has an accuracy of about
$10^{-7}\,\mathrm{cm}^{-1}$. More substantial is the difference in the
spectrum near the dissociation limit. Here the DVR spectrum is perturbed by a
\textquotedblleft ghost\textquotedblright\ state $v=168$. Because of the ghost
state, the resulting number of bound states in the DVR spectrum is incorrect.
The spectral position of the \textquotedblleft ghost\textquotedblright\ state
varies as the parameters of the grid change; for example, the bound spectrum
is no longer perturbed in case of the larger grid (b). By contrast, the
B-spline set spectrum is free of the ghost states regardless of the choice of
the grid.

As we shift to the denser grids (case (b)), the numerical accuracy
of both methods improves. Because of the improved accuracy, in
Table~\ref{Tab:Morse} we list deviations of the numerical energies from the
analytical values. Again, we observe that the B-spline method outperforms the
DVR method in terms of accuracy. This conclusion seem to hold irrespective of
a particular choice of grid. The accuracy of computing the energy of the last
bound level requires special consideration. The relevant outer classical
turning point is located at $R=50.6\,a_{0}$. However, the wavefunction
substantially extends into the classically forbidden region. The small grid
($R_{\mathrm{max}}=100\,a_{0}$) can not fully accommodate this tail. As the size of the cavity
is increased to $2000\,a_{0}$ for the large grid (b), the
B-spline method starts to recover 4-5 significant figures of the exact result for the
energy of the last bound state. Yet the DVR method reproduces only the leading
significant figure.

The superior performance of the B-spline method seems to be due to the compactness of
B-splines. A given B-spline extends only over $k$ intervals of the grid: the B-spline
number $i$ vanishes identically outside a support interval $(t_i, t_{i+k})$.
In particular, it means that for a given coordinate $R$ only a sum of $k$
(in our case $k=15$) B-splines contributes. This is in a stark contrast to the DVR method: here all $N \sim 1000$ rapidly oscillating functions contribute to a value of the wavefunction at a given coordinate, leading to the deterioration of numerical accuracy.
Moreover, it is intuitively clear that while  the DVR sin basis is natural for
describing rapid oscillations in the classically-allowed region, the forbidden region
with its extended exponential tail requires well-balanced interference of  many
basis functions. The accurate description of the classically-forbidden region
becomes more important as we approach the dissociation limit. Namely in this limit
the advantages of using B-splines become more substantial.

\begin{table*}
\begin{center}%
\begin{tabular}
[c]{llllll}\hline\hline
$v$ & Analytical & $E_{v}$, DVR$^{a}$ & $E_{v}$, B-splines$^{a}$ & $\Delta
E_{v}$, DVR$^{b}$ & $\Delta E_{v}$ B-splines$^{b}$\\\hline
162 & $-8.2264075 $ & $-8.2263504$ & $-8.2264050 $ & $9\times10^{-6}$ &
$5\times10^{-13}$\\
163 & $-6.3205792 $ & $-6.3205024$ & $-6.3205766 $ & $1\times10^{-5}$ &
$-1\times10^{-12}$\\
164 & $-4.6655178 $ & $-4.6654299$ & $-4.6655114 $ & $1\times10^{-5}$ &
$1\times10^{-12}$\\
165 & $-3.2612233 $ & $-3.2611065$ & $-3.2612136 $ & $2\times10^{-5}$ &
$8\times10^{-11}$\\
166 & $-2.1076957 $ & $-2.1075487$ & $-2.1074619 $ & $2\times10^{-5}$ &
$9\times10^{-9} $\\
167 & $-1.2049349 $ & $-1.2046878$ & $-1.2023402 $ & $4\times10^{-5}$ &
$2\times10^{-6} $\\
168 & $-0.5529410 $ & $-1.0604859$ & $-0.54813941$ & $6\times10^{-5}$ &
$9\times10^{-6} $\\
169 & $-0.15171398$ & $-0.5527958$ & $-0.14926508$ & $2\times10^{-4}$ &
$4\times10^{-6} $\\
170 & $-1.2538365\times10^{-3}$ & $-0.1507383$ & $-1.0319455\times10^{-3}$ &
$2\times10^{-4} $ & $9\times10^{-8}$\\\hline\hline
\end{tabular}
\caption{ Comparison of the accuracy of the DVR and B-spline methods
in the case of the Morse potential for  two choices of radial grids. Results marked
(a) are for the case of a coarse grid and results marked (b) are for a finer grid.
\label{Tab:Morse}}
\end{center}
\end{table*}

\subsection{Attractive $1/R^{3}$ interactions}
Compared to the Morse potential, realistic molecular potential { {display a long range tail}} leading to a dense vibrational spectrum near the dissociation limit.
The long-range neutral-atom interactions depend on the internuclear distance
as $-C_{n}/R^{n}$, with $n\geq3$.

The most challenging is the case of two atoms interacting via attractive
$-C_{3}/R^{3}$, $C_3>0$, interactions. Such potentials, for example, do not possess
scattering length~\cite{Sha72}.
As a particular example, we consider the {{$A^{1}\Sigma_{u}^+$}}
potential of $^{87}$Rb$_{2}$ dimer correlating to the $5s+5p$ asymptotic
limit, shown in the upper panel of Fig.~\ref{Fig:C3}. This potential is
attractive at large internuclear distances, $V\left(  R\right)  \approx
-C_{3}/R^{3}$. In our specific case $C_3 \approx 17.81$ a.u..

As shown by Le Roy and Bernstein~\cite{LeRBer70} for long-range potentials
varying as $V\left(  R\right)  \approx-C_{3}/R^{3}$%
\begin{equation}
E_{v}=D-\left[  H_{3}~\left(  v_{D}-v\right)  \right]  ^{6},
\label{Eq:LBC3}
\end{equation}
where the constant $H_{3}$ is related to the long-range constant. In our case
the dissociation limit $D=0$.

We plot our computed dependence of $(-E_v)^{1/6}$ on the vibrational quantum number
in the lower panel of Fig.~\ref{Fig:C3}.
We see that the Le Roy-Bernstein formula, Eq.~(\ref{Eq:LBC3}), is followed up to $v \approx 435$.
This equation was derived using semi-classical arguments and
it is known to be violated for the last vibrational
{{levels}} ~\cite{BoiAudVig00}. However, in
our case the deviation from  Eq.~(\ref{Eq:LBC3}) for levels of $v>435$  is simply
due to limitations of the double precision arithmetics (15 significant
figures) used in the computations. Indeed, the energy spectrum spans 14
orders of magnitude: the lowest vibrational state has an energy of
$-2.9 \times 10^{-2}$ a.u., while $E_{v=435}\approx-3.8\times10^{-16}$.
{{Both the B-spline  and the DVR methods}, since they reproduce the entire spectrum in one
shot, do not cope well with the loss of numerical accuracy.
If desired,
numerical accuracy could be improved by switching to quadruple precision
arithmetics.

We find that the B-spline results for levels $v<435$ were insensitive to a particular
choice of the grid, as long as the $R_{\mathrm{max}}$ was well beyond the outer classical
turning point of the wavefunction. By contrast, the DVR code has produced a
multitude of ghost levels, and, for the best choice of the grid parameters, we
were able to reproduce positions of at most 430 vibrational levels.

The computed wavefunction of the $v=434$ level is plotted in Fig.~\ref{Fig:v434C3}.
For this state, the classical turning point is located at $1.9 \times 10^5$ bohr.
The B-spline code was run using the mapped grid with the following parameters:
$R_\mathrm{min} =5.0\, a_{0}$,
$R_\mathrm{max}= 1 \times10^{7}\, a_{0}$, $\beta=0.5$, $E_\mathrm{max}=10^{-15}$. This
corresponds to 1292 grid-points. Notice that the $v=434$  wavefunction has 434 nodes,
yet it was accurately computed using only 1292
grid-points. This is an excellent demonstration of the efficiency of the
mapped-grid technique coupled
with the B-spline method.

\begin{figure}[h]
\begin{center}
\includegraphics[scale=0.75]{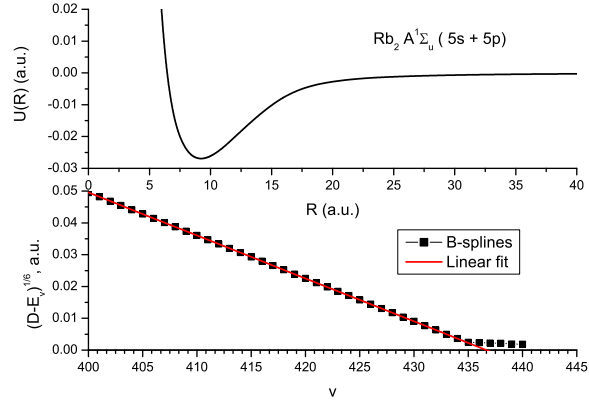}
\end{center}
\caption{Upper panel: Molecular potential $A^{1}\Sigma_{u}^+$ of Rb$_2$ molecule. Lower panel:
comparison of the Le Roy-Bernstein fit (solid line) with the  results
obtained with the B-spline code (squares). }%
\label{Fig:C3}%
\end{figure}

\begin{figure}[h]
\begin{center}
\includegraphics[scale=0.5]{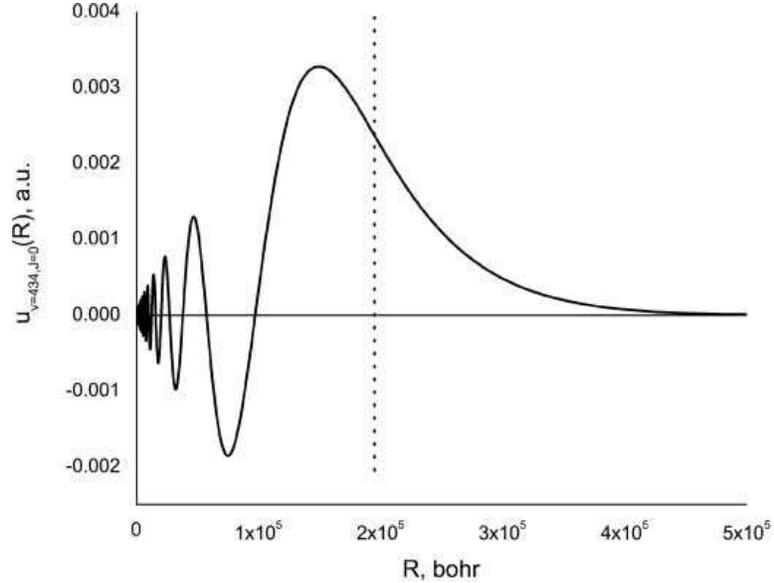}
\end{center}
\caption{Vibrational wavefunction of $v=434, J=0$ level of the Rb $A^{1}\Sigma_{u}^+$
electronic potential as computed in the B-spline method.
The vertical line marks the position of the classical turning
point.}%
\label{Fig:v434C3}%
\end{figure}

\section{Relation between the position of the last bound level and the scattering length}
\label{Sec:eff-range}
Here we consider two atoms interacting at long-range
separations
via attractive $-C_{6}/R^{6}$, $C_6>0$, potentials.
We will employ two scaling
parameters:  the van der Waals length
$\bar{r}_{6}=\left( 2 \mu C_{6}\right)  ^{1/4}$ and energy $\bar{E}%
_{6}=1/\left(  \mu~\left(  \bar{r}_{6}\right)  ^{2}\right)$.
In particular, the regime of  quantum halo states is reached
when the energy of the last bound state is
$-E_{-1} \ll  \bar{E}_{6}$ and { {its spatial extension reaches}} distances
much larger than $\bar{r}_{6}$.

We have investigated the performance of the B-spline method in the case of a
realistic molecular  potential that follows the $1/R^6$ power law at large distances
(this is the case of the ground state of the alkali dimers). The numerical results are
quite similar to the already presented cases of the Morse and $1/R^3$ long-range potential.
Instead, in this section we use the developed method to study universal
relation between the scattering length and the position of the last bound state
in the molecular potential. To this end we focus on a simple model of hard-core sphere with the van der Waals tail. In this model
the short-range physics is modeled
 by placing an impenetrable wall at $R=R_{0}$:
\begin{equation}
V(R) =\left\{
\begin{array}
[c]{cc}%
\infty \, , & R< R_0\\
 -C_{6}/R^{6} \, , & R\ge R_0
\end{array}
\right. \, .
\end{equation}

This simple model offers insights
into the universal laws of low-energy scattering.
Let us enumerate
several analytical results~\cite{GriFla93,Gao04a} for this model relevant to
our discussion. These are formulated in terms of the {{scaling factor}}
\[
   \bar{a} = \frac{2 \pi}{(\Gamma(1/4))^2} \bar{r}_{6} \approx 0.477989 \, \bar{r}_{6},
\]
and accumulated phase inside the potential
$$
 \Phi = \frac{\bar{r}_{6}^2}{2 R_0^2}\, .
$$
{{which determines the physics close to threshold}}.
Indeed, the number of bound states is given by~\cite{GriFla93}
$$
   N_b =  \lfloor \Phi/\pi -7/8  \rfloor +1 \, ,
$$
and the scattering length $a$ by~\cite{GriFla93}
\begin{equation}
 a = \bar{a} \left( 1 - \tan( \Phi - 3\pi/8) \right) \, .
\label{Eq:GFscatLen}
\end{equation}
{ {For $|a|/r_{6} \gg 1$, there is either a bound level close to the dissociation limit ($a>0$) or a virtual state ($a<0$).}}

In our numerical study, we take $C_{6}=6851$ for the ground-state Cs dimer
~\cite{DerJohSaf99}, and a reduced mass for $^{133}$Cs atoms.
For $^{133}$Cs$_{2}$
molecule $\bar{r}_{6}\approx 202~a_{0}$,
and $\bar{E}_{6} \approx 4.4 \times 10^{-5}~\mathrm{cm}^{-1}$.
Increasing $R_0$, the position of the inner ``hard'' wall of the potential,
reduces the number of bound states in the potential. For example,
we find from analytical formula that a new bound state
appears at the value of $R^*_0 \approx 6.02073$  a$_0$. The potential
binds 180 states for $R_0$ just below $R^*_0$ and 179 states for
$R_0$ just above $R^*_0$.

For our initial numerical test we choose the position of the inner wall at $R_{0}=6.02$
bohr. B-spline method reliably produces all
179 bound states and reveals a loosely bound state with the energy of
 $-9.33 \times 10^{-12}$ a.u. We verified that the energies of the states
 near the dissociation limit follow the Le Roy-{{Bernstein}} pattern
 (similar to the analysis presented in Fig.~\ref{Fig:C3} for the $1/R^3$ potential.)
In this case, however, some additional observations can be made.

For  $R_{0}=6.02$ a$_0$, the scattering length, Eq.~(\ref{Eq:GFscatLen})
is large and positive, $a=+796$  a$_0$.
Large and positive scattering lengths result from having a bound state
just below the threshold. In this regime,
the energy of the last bound state may be approximated by~\cite{Gao04a}
\begin{eqnarray}
 E_{-1}^\mathrm{QDT} &\approx& -\frac{1}{2\mu} \frac{1}{(a - \bar{a})^2} \times \nonumber\\
 &&\left( 1 +
 c_1   \frac{\bar{r}_{6}}{ (a - \bar{a}) } +
 c_2   \frac{\bar{r}_{6}^2}{ (a - \bar{a})^2 }
 \right) \, , \label{Eq:GaoLast}
\end{eqnarray}
where $c_1 \approx 0.4387552 $, $c_2 \approx -0.2163139$.
The above expression was derived using the quantum defect theory and
it substantially differs from the commonly-used effective
range expansion formula
\begin{equation}
 E_{-1}^\mathrm{eff} = - \frac{1}{2\mu} \frac{1}{a^2} \, .
\label{Eq:effRandgeLast}
\end{equation}
From  Eq.~(\ref{Eq:GaoLast}), we find
$E_{-1}^\mathrm{QDT} \approx -9.35 \times 10^{-12}$, while
the effective-range formula results in
$E_{-1}^\mathrm{eff} \approx -6.52 \times 10^{-12}$. Clearly,
our numerical result,  $-9.33 \times 10^{-12}$ a.u., supports the
analytical analysis\cite{Gao04a}. In this calculation, the parameters of the grid were chosen to be
$R_{\min}=R_{0}$, $R_{\max}=5 \times 10^{4},\beta=0.4$, with the number of points
2407. When the number of points was reduced by a factor of 3, the energy
of the last bound state was affected in the third significant figure.
We again notice that the DVR method was unable to match the numerical
accuracy of the B-spline approach.

While offering an improved accuracy over the effective-range
expression, the QDT Eq.(\ref{Eq:GaoLast}) is still an approximate result.
In Fig.~\ref{Fig:GaoPlot},
we compare the QDT prediction with our numerical results. Here we move
the position of the inner wall just below the {{critical}} value of $R^*_0 \approx 6.02073$,
at which the least bound state disappears.
The range of the values
for the position of the inner wall was chosen so that the scattering length remained
positive. An increase in $R_0$  translates into
increasingly larger values of the scattering length.
For $a/\bar{a} \gg 1$, i.e., near the threshold,
both the effective-range and the QDT  results become
identical. As the scattering length decreases, the
effective range approximation rapidly loses
its accuracy. Our comparison in
Fig.~\ref{Fig:GaoPlot} clearly demonstrates that,
compared to the conventional effective-range theory, the QDT
approximation is applicable over a much wider range of parameters.
At the same time,
as $R_0$ is decreased from its critical value,
the QDT approximation starts to break down at $R_0 \approx 6.017$ bohr.
The relevant parameter governing the validity of  Eq.(\ref{Eq:GaoLast})
is the { reduced} scattering length $a/\bar{a}$: { {the critical value}}
$R_0 \approx 6.017$  a$_0$ corresponds to  $a/\bar{a} \approx 2$.
To reiterate, the QDT formula, Eq.(\ref{Eq:GaoLast}), is an excellent approximation
as long as  $a/\bar{a} > 2$, while the effective range approximation
requires $a/\bar{a} \gg 1$.

Finally, it is worth pointing out
that  our method is robust enough to reproduce
halo states of diatomic molecules bound by the van der Waals forces.
We varied $R_0$ just below the threshold value
and examined the energies of the least bound state produced by the B-spline method.
For example, for $R_0=6.0207$, we obtain with the B-spline
code $E_{-1} = -1.71 \times 10^{-14}$ a.u.,
while analytical results are
$E_{-1}^\mathrm{QDT} \approx -1.82 \times 10^{-14}$ a.u. and
$E_{-1}^\mathrm{eff} \approx -1.78 \times 10^{-14}$ a.u.
The binding energies are four orders of magnitude smaller than the van der Waals
energy. At the same time, the corresponding scattering length,
governing the extent of the wavefunction, is about $2 \times 10^4$ bohr,
i.e., two orders of magnitude larger than the van der Waals length.
Satisfying both enumerated conditions signifies reaching the universal
regime of quantum halo states.

\begin{figure}[h]
\begin{center}
\includegraphics[scale=0.85]{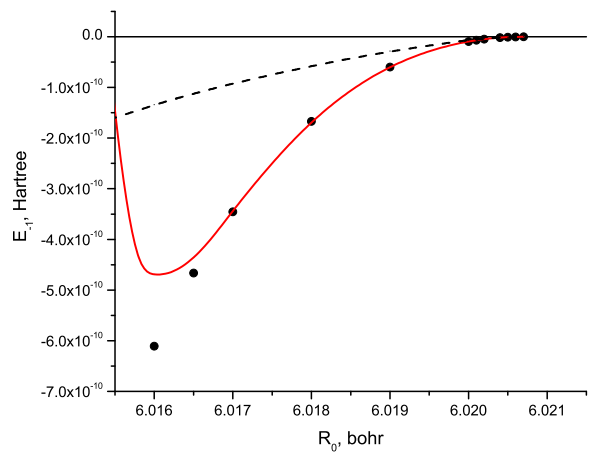}
\end{center}
\caption{Energy of the last bound state as a function of the position
of the inner wall of the model potential. Dots mark numerical results
obtained with the B-spline approach. Predictions of
the effective-range approximation
are shown with a dashed line and that of  the quantum-defect theory --
with a solid line.}%
\label{Fig:GaoPlot}%
\end{figure}

\section{Conclusion}
\label{Sec:Conclusion}
With the experimental control of quantum-mechanical systems becoming
more refined, new theoretical tools have to be adopted to meet the new
challenges. Recently, fragile Feshbach (quantum halo-state) molecules
became an experimental reality (see e.g., Ref.~\cite{MarFerKoo07}). Motivated by this progress, here
we developed a numerical method for solving the Schr\"{o}dinger equation
for diatomic molecules based on the B-spline finite basis sets.
The method produces a numerically complete quasi-spectrum of rovibrational states.
We find, that  B-splines offer an
accurate description of the loosely-bound molecular states near the dissociation
limit. The quasi-spectrum is entirely devoid of the unphysical ghost states { {which appear in DVR method and require special effort to be eliminated \cite{WilDulMas04,kallush2006}}} .
Moreover, coupled with the ``mapped grid'' method of Ref.~\cite{WilDulMas04},
the representation is both accurate and efficient: both rapidly-oscillating
part of the wavefunction in the classically-allowed region and the slowly-varying
exponential tail in the  classically-forbidden region are adequately reproduced.
As an application of the developed method we investigated
the universal
law relating the energy of the last bound state to the scattering length.
We find that the new QDT formulation of such a law for $1/R^6$ potentials
by Gao~\cite{Gao04a} remains valid over a substantially wider range of parameters than
the commonly-used effective-range approximation.

\section*{Acknowledgements}
AD would like to thank Walter Johnson for introduction to B-splines and also Laboratoire Aime Cotton for hospitality during a visit when a part of this work was carried out. The work of AD was supported in part by US National Science Foundation grant No. PHY-06-53392 and in part
by the National Aeronautics and Space Administration under Grant/ Cooperative Agreement No. NNX07AT65A issued by the Nevada NASA EPSCoR program.


\end{document}